\documentclass[article]{aa}
\usepackage{epsfig}
\begin{document}

\newcommand{\gsim}{\hbox{\rlap{$^>$}$_\sim$}}
  \thesaurus{06;  19.63.1}
\authorrunning{S. Dado, A. Dar \& A. De R\'ujula}
\titlerunning{The supernova associated with GRB 020405 }
\title{The supernova associated with GRB 020405 }
\author{Shlomo Dado$^{^1}$, Arnon Dar$^{^1}$ and
A. De R\'ujula$^{^2}$}
\institute{1. Physics Department and Space Research Institute, Technion,
               Haifa 32000, Israel\\
           2. Theory Division, CERN, CH-1211 Geneva 23, Switzerland}
\maketitle

\begin{abstract}

We use the very simple and successful Cannonball (CB) model  of gamma ray
bursts (GRBs) and their afterglows (AGs) to analyze the observations of
the mildly extinct optical AG of the relatively nearby GRB 020405.
We show that GRB 020405 was associated with a
1998bw-like supernova (SN) at
the GRB's redshift that appeared dimmer and redder than SN1998bw
because of extinction in the host and our Galaxy.
The case for the SN/GRB association
---advocated in the CB model--- is becoming indubitable. We discuss the 
extent to which the GRB/SN connection is model-dependent.

\end{abstract}


\section{Introduction}
In the Cannonball Model of GRBs (Dar and De R\'ujula 2000, briefly
reviewed in De R\'ujula 2002a,b) long
duration GRBs are produced by highly relativistic jetted plasmoids
(cannonballs) in core-collapse supernovae akin to SN1998bw
(Dar and Plaga 1999; Dar 1999a; Dar and De R\'ujula 2000 and references
therein).  Possible evidence for an SN1998bw-like contribution to a GRB
afterglow (Dar 1999a; Castro-Tirado \& Gorosabel 1999) was first reported by
Bloom et al.~(1999) for GRB 980326, but its unknown redshift prevented a
categorical conclusion. The AG of GRB 970228 (at a redshift $\rm
z=0.695$) appears to be overtaken by a light curve akin to that of
SN1998bw (at $\rm z_{bw}=0.0085$), when properly scaled by their
differing redshifts (Dar 1999b; Reichart 1999; Galama et al.~2000).
Evidence of a similar associations was found for GRB 000418 (Dar
and De R\'ujula 2000; Dado et al. 2002a), GRB 980703 (Holland et
al.~2000), GRB 991208 (Castro-Tirado et al.~2001), GRB 990712 (Bjornsson
et al.~2001), GRB 970508 (Sokolov et al. 2001), GRB 000911 (Lazzati et al.
2001; Dado et al. 2002b), GRB 010921 (Dado et al. 2002c) and GRB 011121
(Bloom et al. 2002a; Dado et al. 2002d).
In the CB model, even the very nearby GRB 980425 and its
associated SN (1998bw) are ``normal'': only the low redshift and
relatively large viewing angle are unique
(Dar and De R\'ujula 2000, Dado et al. 2002e).

Unlike supernovae of type Ia (SNe Ia), core-collapse supernovae (SNe
II/Ib/Ic) are far from being standard candles. But if their ejecta are
fairly asymmetric ---as they would be if a fair fraction of them emit two
opposite jets of cannonballs--- much of the diversity could be due
to the varying angles from which we see their non-spherically
expanding shells. Exploiting this possibility to its extreme, i.e., using
SN1998bw as an ansatz standard candle, Dar and De R\'ujula (2000) and Dado et
al. (2002a)  have shown that the optical AG of {\it
all} relatively nearby GRBs with known redshift (all GRBs with $\rm
z<1.12$) contain evidence or clear hints for an SN1998bw-like contribution
to their optical AG, suggesting that most ---and perhaps all--- of the
long duration GRBs are associated with 1998bw-like supernovae (in the more
distant GRBs, the ansatz standard candle could not be seen, and it was not
seen).  In several of the above cases, however, scarcity of data,
lack of spectral
information and multicolour photometry and the uncertain extinction in the
host galaxy prevented a firm conclusion. Thus, every new instance is still
interesting, it might take a few more clear cases to reach a generally
accepted conclusion.

On 2002 April 5.028773 UT the long duration ($\sim 40$s) GRB 020405 was
detected and localized by Ulysses, Mars Odyssey - HEND, and BeppoSAX
(Hurley et al.  2002).  Its optical AG was first detected in the R-band,
17.5 h after the burst (Price et al. 2002a) and its fading
was followed in the I, R, V and B bands (Castro-Tirado et al.  2002;
Palazzi et al. 2002; Hjorth et al. 2002a,b;  Price et al. 2002b; Gal-Yam et
al. 2002;  Covino et al. 2002a,b,c; Bersier et al. 2002).
Its redshift, $\rm z=0.69$, was determined
(Masetti et al. 2002, Price et al. 2002c) from emission lines of its
likely host galaxy.  The optical AG was well fitted
by a $\rm t^{-1.72}$ power-law decay, but the late time R-band measurements
with the Magellanic 6.5m Baade telescope, on April 18th, and with the  
100'' du
Pont telescope at Las Campanas, on May 3rd, lie significantly above
the power-law extrapolation  (Bersier et al. 2002).

The late time AG of GRB 020405 was observed with HST at
several epochs spanning 19-31 days after the GRB and, to remove the
host contribution,  two months after the GRB (Price et al. 2002d).
An excess of flux was found in the
HST images, compared to an extrapolation of the light-curve from early
times and was identified as a SN associated with GRB 020405, redder  
than a SN1998bw displaced to
$\rm z=0.69$, and dimmer than this ansatz by
about one half magnitude (Price et al. 2002d).

In this letter we use the Cannonball Model to estimate the extinction in
the host galaxy of GRB 020405, and to predict the late time optical AG. We
show that the evidence from HST is clear:  GRB 020405 was indeed
associated with a standard-candle 1998bw-like supernova at $\rm z=0.69$,
appearing somewhat dimmer and redder just because of  the
extinction in the host galaxy and in ours.

\section{The CB model}

In the CB model, long-duration GRBs and their AGs are produced in core   
collapse SNe by jets of highly relativistic ``cannonballs'' that
pierce through the SN shell.
Crossing this shell with a large Lorentz factor $\gamma$, the
surface of a CB is collisionally heated to keV temperatures and the
radiation it emits when it reaches the transparent outskirts of the
shell ---boosted and collimated by the CB's motion---
is a single $\gamma$-ray pulse in a GRB. The cadence
of pulses reflects the chaotic accretion processes and is not
predictable, but the individual-pulse temporal and spectral properties are.
A long list of general properties
(Dar and De R\'ujula 2001) of GRB pulses 
is reproduced in the CB-model, in which, unlike
in the standard ``fireball'' models (for a review, see, e.g. Ghisellini 2001)
 the GRBs' $\gamma$'s have a thermal ---as opposed to synchrotron--- origin.

A CB exiting a SN shell soon becomes transparent to its own enclosed
radiation. At that point, it is still expanding and cooling adiabatically
and by bremsstrahlung. The bremss spectrum is hard and dominates
the very early X-ray AG (for a minute or two)  
with a fluence of predictable magnitude decreasing
with time as $t^{-5}$ until synchrotron emission from the ISM electrons 
that enter the CB takes over. All X-ray 
AGs are compatible in magnitude and shape with this prediction
(Dar and De R\'ujula 2002a) for the very early AG,
for which, as far as we know, there is no
standard (fireball-model) counterpart.  The
optical AGs for which the data also start very early after the GRB
show an early decline $\propto t^{-2}$. This initial decline is produced
when the CB is plowing through a $\sim r^{-2}$
density-profile of the ``wind'' from the  SN progenitor
(Dado et al. 2002a).
In the fireball models, the absence of ``windy'' signatures is a problem: 
of the score of observed cases, only one has ``clear evidence for a
wind-fed circumburst medium'' (Price et al. 2002).

In the CB model the AGs observed at later times
have three origins: the ejected CBs, the
concomitant SN explosion, and the host galaxy. These components are
usually unresolved in the measured ``GRB afterglows'', so that the
corresponding light curves and spectra are the cumulative energy flux   
density $\rm    F_{AG}=F_{CBs}+F_{SN}+F_{HG}$.
The contribution of the  host galaxy (HG) is usually determined
by late time observations when the CB and SN contributions become
negligible.

Let the energy flux
density of SN1998bw at redshift $\rm z_{bw}=0.0085$ (Galama et al. 1998)   
be $\rm F_{bw}[\nu,t]$. For a similar SN placed at a redshift $\rm z$:   
\begin{eqnarray}
{\rm F_{SN}[\nu,t] = } && {\rm{1+z \over 1+z_{bw}}\;
{D_L^2(z_{bw})\over D_L^2(z)}}\, \times\nonumber \\ &&
{\rm
F_{bw}\left[\nu\,{1+z \over 1+z_{bw}},\;t\, {1+z_{bw} \over 1+z}\right]\;   
A(\nu,z)}\, ,
\label{bw}
\end{eqnarray}
 where $\rm A(\nu,z)$ is the attenuation along the line
of sight and $\rm D_L(z)$ is the
luminosity distance (we use a cosmology with
${\rm \Omega_M}=0.3$, ${\rm
\Omega_\Lambda}=0.7$ and  $\rm H_0=65$ km/s/Mpc).

In its rest frame the optical AG of a CB  
is given by:
\begin{equation}
\rm F_{_{CB}}[\nu,t]=
{f\, [\gamma(t)]^2\over \nu_b}{[\nu/\nu_b]^{-1/2}\over
\sqrt{1+[\nu/\nu_b]^{(p-1)}}}\; ,
\label{fluxdensity2}
\end{equation}
where $\rm f$ is a normalization constant (see Dado et al. 2002e for its
theoretical estimate), $\rm \gamma(t)$ is the Lorentz
factor of the CB, $\rm p\approx 2.2 $ is the spectral index of the radiating 
electrons in the CB  and $\rm \nu_b$ is the ``injection bend'' frequency.
For an interstellar density $\rm n_p$:
\begin{equation}
\rm \nu_b \simeq 1.87\times 10^3\, [\gamma(t)]^3\,
\left[{n_p\over 10^{-3}\;cm^3}\right]^{1/2}\, Hz.
\label{nubend}
\end{equation}
The theoretical motivation, as well as the excellent observational support
for this ``bend'', are discussed in Dado et al. (2002e).
An observer in the GRB progenitor's rest system,
viewing a CB at an angle $\theta$, sees its radiation
Doppler-boosted by a factor $\delta$:
\begin{equation}
\rm \delta(t)\equiv
{1\over\gamma(t)\,(1-\beta(t)\cos\theta)}
\simeq {2\,\gamma(t)\over 1+\theta^2\gamma(t)^2}\; ,
\label{doppler}
\end{equation}
where the approximation is valid in the domain of interest for GRBs:
large $\gamma$ and small $\theta$.
The cannonballs' AG spectral energy density $\rm F^{obs}_{CB}$
seen by a cosmological observer at a redshift $\rm z$
(Dar and De R\'ujula, 2000), is:
\begin{equation}
\rm F^{obs}_{CB}[\nu,t]\simeq
\rm {A(\nu,z)\, (1+z)\,\delta(t)^3
                    \over 4\, \pi\, D_L^2}\,
F_{_{CB}}\left[{(1+z)\,\nu\over\delta(t)},{\delta(t)\,t\over 1+z}
\right]\! .
\label{Fnuobser}
\end{equation}

For an interstellar medium of constant baryon density $\rm n_p$, the   
Lorentz factor  $\rm\gamma(t)$ is given by (Dado et al. 2002a):
\begin{eqnarray}
\rm \gamma&=&\rm\gamma(\gamma_0,\theta,x_\infty;t)
=\rm {B^{-1}} \,\left[\theta^2+C\,\theta^4+{1/C}\right]\nonumber\\
\rm C&\equiv&\rm
\left[{2/  
\left(B^2+2\,\theta^6+B\,\sqrt{B^2+4\,\theta^6}\right)}\right]^{1/3}
\nonumber\\
\rm B&\equiv&\rm
{1/ \gamma_0^3}+{3\,\theta^2/\gamma_0}+
{6\,c\, t/ [(1+z)\, x_\infty]}
\label{cubic}
\end{eqnarray}
where $\gamma_0=\gamma(0)$, and
$\rm x_\infty\equiv N_{CB}/(\pi\, R_{max}^2\, n_p)$
characterizes the CB's slow-down in terms of
$\rm N_{CB}$: its baryon number, and $\rm R_{max}$:
its radius (it takes a distance $\rm x_\infty/\gamma_0$ for
the CB to half its original Lorentz factor).

The selective extinction, $\rm A(\nu,t)$ in Eq.~(\ref{bw}), can be
estimated from the difference between the observed spectral index {\it at
very early time when the CBs are still near the SN} and that expected in
the absence of extinction. Indeed, the CB model predicts ---and the data
confirm with precision--- the gradual evolution of the effective  
optical spectral
index towards the constant value $\approx -1.1$ observed in all
``late'' AGs (Dado et al. 2002a). The ``late'' index
is independent of the attenuation in the host
galaxy, since at $\rm t>1$ (observer's) days after the explosion, the CBs are
typically already moving in the low column density, optically  
transparent halo of
the host galaxy.

The comparison of the predictions of Eq.~(\ref{Fnuobser}) with the
observations of optical and X-ray AG light curves and of broad-band
spectra is discussed in Dado et al. (2002a and 2002e, respectively).
The results ---for {\it all} GRBs of known redshift--- involve a total  
of only five parameters, and are very satisfactory. The CB-model  
results concerning X-ray
lines in GRB AGs  are also
exceptionally predictive, simple and encouraging (Dado et al. 2002f).

\section{GRB 020405 in the CB model}

We have first fitted the CB-model predictions to the
B, V, R and I light curves of the AG of GRB 020405, as
observed during the first 5 days after burst.
In using Eq.~(\ref{fluxdensity2}),
we assumed an electron spectral index $\rm p
= 2.2$, compatible with that of all other GRB AGs (Dado et al. 2002a). The
fitted parameters are: $\rm \gamma_0 = 645$, $\rm \theta =
0.42$ mrad, and $\rm x_\infty = 0.31$ Mpc.
After correcting for selective extinction in our Galaxy ($\rm E(B-V)=0.054$ 
mag towards  GRB 020405;
Schlegel et al. 1998), Bersier et al. (2002) found that the
broad band BVRI spectrum of the AG, 1.3 days after the burst,
had a spectral shape $\rm F_{obs}\sim \nu^{-1.43\pm 0.08}$.
In the CB model, the unextinct spectral index in  optical and
nearby frequencies evolves from $\sim -0.5$ to $\sim -1.1$, as the
injection bend frequency of Eq.~(\ref{nubend}) diminishes
with time. At the time of these observations, our
fit to the AG time-evolution results in a predicted index
$ -0.8\pm 0.1$.  If the difference with the observed index
is due to selective extinction in the host galaxy, then  
$E(B-V)=0.24\pm 0.05$
and the attenuation factors in the
I, R, V and B bands (e.g., Whittet 1992) are, respectively,
$\rm A(\nu,z)\sim 0.58\,, 0.50\,, 0.42~and~ 0.34$. These attenuations,
together with the Galactic extinction, were used to dim the
contribution to the AG
of a SN1998bw-like supernova at the redshift of GRB 020405.
The resulting late-time I, R, V and B light curves
are presented in Fig.~(1).

The agreement between theory and observations in Fig. (1) is
surprisingly good, in view of the large observational uncertainties and
the theoretical approximations. The presence of an SN1998bw-like
signal is completely convincing. With our consistent estimate of
extinction in the host galaxy, the underlying SN is indistinguishable,
within errors, from a standard candle SN1998bw.

\begin{figure}[t]
\vspace{-.5cm}
\hskip 2truecm
\hspace*{-2.1cm}
\epsfig{file=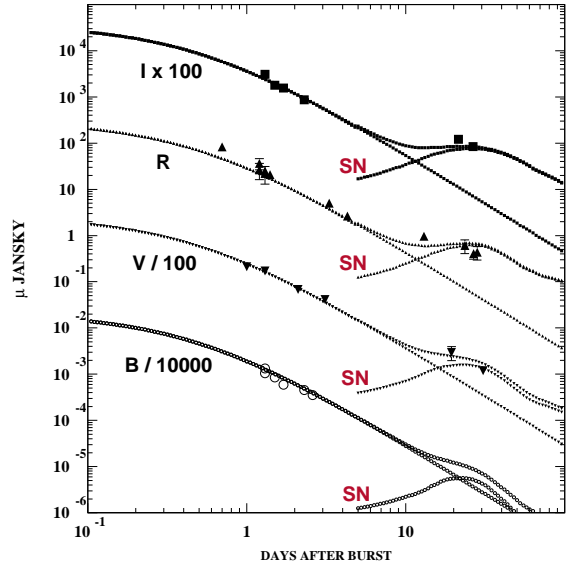, width=8.5cm} \\
\vspace{-0.5cm}
\caption{CB model  fit to the
measured I, R, V, and B-band AG of GRB 020405,
multiplied by 100, 1, 1/100, 1/10000, respectively.
The observations are not corrected
to eliminate the effect of extinction, thus
the theoretical contribution  from a SN1998bw-like supernova,   
Eq.~(\ref{bw}),
was dimmed by the known extinction in the Galaxy and our consistently
estimated extinction in the host. The contribution of the host galaxy,
subtracted from the data by the HST observers, is not included in the fit.}
\label{figr0425}
\end{figure}

\section{The GRB/SN association in various models}

In the CB model, GRBs are associated with SNe.
The model is so simple and predictive that it allowed us
to use, in the case of GRB 011121, the first 2 days of
R-band data to fit the parameters describing the CBs' contribution to  
the AG. These fitted parameters were used 
to predict explicitly the AG evolution 
and that
{\it the SN will tower in all bands over the CB's declining light curve  
 at day $\sim 30$ after burst} (Dado et al. 2001). The comparison
with the data, gathered later, vindicated the presence of a  
SN1998bw-like signal
(Dado et al. 2002d), as in GRB 020405 and
all the other cases for which the SN was, in practice,
detectable (Dado et al. 2002a). To what extent is the GRB/SN
association model-dependent?

There are other models that link
long-duration GRBs with SN explosions: the hypernova model
(Paczynski 1998), the collapsar model (Woosley 1993) the supranova
model (Vietri and Stella 1998) and the
currently favoured conical jet models (see, e.g. Rossi et al. 2002, Zhang
and Meszaros 2002,  Salmonson and Galama 2002, Granot et al. 2002),
in which the authors depart from
the earlier view that it is a good approximation (Rhoades, 1999)
to consider conical jets of various {\it opening} angles, pointing
exactly to the observer, and having a uniform Lorentz factor
(in the CB model the {\it viewing} angle is the {\it only one} that matters,
and there is no need to introduce ``jet profiles'' of varying rapidity).
A common feature of all these models is that they
are not sufficiently specific to repeat the CB-model exercise of the
previous paragraph. In the jetted models, for instance,
the AGs (unlike the smoothly-varying
data) have ``breaks''. The early data on GRB 011121
do not tell where the break ``is'': they cannot be extrapolated.

In the CB model, the exceptionally close-by GRB 980425
($\rm z=0.0085$) and its associated
SN1998bw are {\bf not} intrinsically exceptional. Because it was viewed at
an exceptionally large angle ($\sim 8$ mrad), the GRB's
$\gamma$-ray fluence was comparable to that of more distant
ones, viewed at  $\theta\sim 1$ mrad. That is why its optical AG
was dominated by the SN, except, perhaps, for the last measured point
(Dar and De R\'ujula 2000). The X-ray AG
is also of ``normal'' magnitude, it is {\bf not} emitted by the SN;
its fitted parameters allowed us to predict successfully the
magnitude of the cited last optical point (Dado et al. 2002a).
The normalization, time and frequency
dependence of the radio AG of this GRB are also ``normal'', and due to
the CB, not the SN (Dado et al. 2002e). SN1998bw, deprived of its
``abnormal'' X-ray and radio emissions (which it did not emit!), loses  
most of
its ``peculiarity''. Given that GRB 980425 and SN 1998bw are not
exceptional, the use of this SN as a standard candle
to look for in other AGs (Dar 1999b) is, in the CB model, entirely natural.
The surprise is how well this ansatz works (Dado et al. 2002a).

In the hypernova and collapsar models the parent star is supposed
to be exceptionally massive. In the supranova model, it must
have exploded months before the GRB emission. Moreover, in these
models, and in the currently favoured jetted models, GRB 980425
stands in a class by itself.  Thus, there is
every reason {\bf not} to expect its associated SN (1998bw) to be similar
to the SNe associated with other GRBs. In the CB model ---and in the data---
the opposite is true.

\section{Conclusion}

In Dar and De R\'ujula (2000) we argued that long-duration
GRBs may all be associated with 1998bw-like supernovae,
and that the apparent variability of core-collapse SNe may to a large
extent be due to a spread of viewing angles, relative to
the CB-emission axis. In Dado et al. (2002a) we showed how
surprisingly successful the ansatz of an associated supernova
identical to 1998bw was, when confronted with the observations
for optical and X-ray AGs. The AGs of some GRBs discovered after
these quoted works ---GRB 000911 (Dado
et al. 2002b), GRB 010921 (Dado
et al. 2002c), GRB 011121 (Dado et al.
2002d) and GRB 020405, discussed here--- strengthen the
conclusion: so far, in
all AGs in which a SN like SN1998bw could be seen (in practice, in the cases
with redshift $\rm z< 1.12)$, it was seen, and it was
compatible in magnitude and colour with an  SN1998bw
standard-candle!

It goes without saying
that there are no standard candles. It is just that the current data are not 
precise enough to detect significant deviations.
But the important fact is
that the SN1998bw-like supernovae allegedly associated with
all long-duration GRBs (Dado et al. 2002a, and references therein)
{\bf happen to be there}.

{\bf Acknowledgment:} This research was supported in part by the Helen 
Asher Space Research Fund and by the VPR fund for research at the Technion.
We thank the referee Brian McBreen for useful suggestions.

\end{document}